\documentclass{jnmp}

\usepackage{amsmath}
\usepackage{cite}
\JNMPnumberwithin{equation}{section}

\begin{document}

\renewcommand{\evenhead}{Extended Prelle-Singer Method}
\renewcommand{\oddhead}{V K Chandrasekar, M Senthilvelan and M Lakshmanan}

\thispagestyle{empty}

\copyrightnote{2004}{M.~Lakshmanan}

\Name{Extended Prelle-Singer Method and Integrability/Solvability of a Class of
 Nonlinear $n$th Order Ordinary Differential Equations
 \footnote{Article dedicated to Professor F. Calogero on his $70^{th}$ birthday}}

\label{firstpage}

\Author{V K Chandrasekar, M Senthilvelan and M Lakshmanan}

\Address{Centre for Nonlinear Dynamics, Bharathidasan University \\
Tiruchirapalli - 620 024, India \\
~~E-mail: sekar@cnld.bdu.ac.in, senthilvelan@cnld.bdu.ac.in, 
lakshman@cnld.bdu.ac.in 
\\}

\Date{\today}

\begin{abstract}
\noindent
We discuss a method of solving $n^{th}$ order scalar ordinary differential
equations by extending the ideas based on the Prelle-Singer (PS) procedure for
second order ordinary differential equations. We
also introduce a novel way of generating additional integrals of motion from a single
integral. We illustrate the theory for both second and third order equations
with suitable examples. Further, we extend the method to two coupled second
order equations and apply the theory to two-dimensional Kepler problem and
deduce the constants of motion including Runge-Lenz integral.
\end{abstract}

%
\section{Introduction}
During the past two decades or so immense interest has been shown in identifying
integrable/solvable nonlinear dynamical systems. Different ideas
have been developed or employed in isolating such nonlinear ordinary/partial
differential equations \cite{Lakshmanan}. Focussing our attention on ordinary differential
equations (ODEs), one finds that using certain novel ideas, Calogero and his 
coworkers have
generated a wide class of integrable/solvable nonlinear systems and explored
their underlying features \cite{Calogero1,Calogero3, Calogero2}. In this paper we 
introduce another straightforward method of identifying
such solvable and/or integrable equations. To do so we employ the extended 
Prelle-Singer (PS) procedure \cite{Prelle}.

Sometime ago Prelle and Singer \cite{Prelle} have proposed a 
procedure for solving first order ODEs that presents the solution in terms 
of elementary functions if such a solution exists.  The attractiveness 
of the PS method is that if the given system of first order ODEs has a solution 
in terms of elementary functions then the method guarantees that this solution 
will be found.  Very recently Duarte et al \cite{Duarte} have modified 
the technique developed by Prelle and Singer \cite{Prelle} and applied 
it to second order ODEs.   
Their approach was based on the conjecture that if an elementary solution 
exists for the given second order ODE then there exists at least one 
elementary first integral $I(t,x,\dot{x})$ whose derivatives are all 
rational functions of $t$, $x$ and $\dot{x}$.  For a class of  
systems these authors have deduced first 
integrals and in some cases for the first time through their procedure 
\cite{Duarte}. Recently
the present authors have generalized the theory given in \cite{Duarte}
and pointed out a
procedure to obtain all the integrals of motion/general solution and solved a 
class of nonlinear equations \cite{Chand1, Chand2, Chand3}. 

The PS procedure has many attractive features. For a large class of integrable
systems, this procedure provides the integrals of motion/general solution in a
straightforward way. In fact this is true for any order. The PS method not only
gives the first integrals but also the underlying integrating factors. Further,
like Lie-symmetry analysis and Noether's theorem the PS method can also used to
solve linear as well as nonlinear ODEs. In addition to the above, the PS
procedure is applicable to deal with both Hamiltonian and non-Hamiltonian
systems.

In this paper we extend the above theory \cite{Duarte, Chand1, Chand2, Chand4} 
to $n^{th}$ order scalar
ODEs and derive a relation which connects integrals of motion with the
integrating factors. We demonstrate the method to second and third order ODEs.
We also introduce a novel way of generating all the integrals of motion from a single
integral, {\it which is applicable to a class of ODEs}, and demonstrate 
our ideas by considering the previous examples. Finally, we prolong the 
method to two coupled second order ODEs. 

We note here that several works have been undertaken earlier to study and
classify systematically the $n$th order ODEs based on different ideas. To name 
a few, besides the works of Calogero \cite{Calogero1, Calogero3}, we cite, 
Lie symmetry approach \cite{Ibragimov1,Ibragimov2,Mahomed, Olver},
the method of Bluman and Anco \cite{Anco, Bluman}, 
Painlev\'e analysis \cite{Lakshmanan, Lakshmanan1} and 
so on.

The organization of the material is as
follows. In the following section we extend the theory of modified PS method to
$n^{th}$ order ODEs. In Sec.3, we describe a procedure to generate second, third
and higher integrals of motion from a single integral, provided the first
integral can be written in certain specific form. In Secs.~4 and 5,
respectively, we
demonstrate the theory to second and third order ODEs with suitable examples. In
Sec. 6, we briefly discuss the application of PS procedure to higher order ODEs.
In
Sec.~7 we prolong the theory to two-coupled second order ODEs and illustrate the
theory with an example, namely the two-dimensional Kepler problem. We give our 
final remarks in Sec.~8.

\section{Prelle-Singer Procedure}
\label{sec2}
Let us consider a class of $n^{th}$ order ODEs of the form 
\begin{eqnarray} 
x^{(n)}=\frac{P}{Q},\qquad  {P,Q}\in C{[t,x,x^{(1)},x^{(2)},...,x^{(n-1)}]}, 
 \label{net1}
\end{eqnarray}
where $x^{(1)}=\frac{dx}{dt},x^{(2)}=\frac{d^2x}{dt^2}$ and $x^{(n)}=\frac{d^nx}{dt^n}$
and $P$ and $Q$ 
are polynomials in $t,x,x^{(1)},x^{(2)},$ $..., x^{(n-2)}$ and $x^{(n-1)}$ with 
coefficients in the 
field of complex numbers. Let us assume that the ODE (\ref{net1}) 
admits a first integral $I(t,x,x^{(1)},x^{(2)},...,x^{(n-1)})=C,$ with $C$ constant on the 
solutions, so that the total differential gives
\begin{eqnarray}  
dI={I_t}{dt}+{I_{x}}{dx}+{I_{x^{(1)}}}{dx^{(1)}}
+\ldots +{I_{x^{(n-1)}}{dx^{(n-1)}}}=0, 
\label{net3}  
\end{eqnarray}
where subscript denotes partial differentiation with respect 
to that variable. Rewriting equation~(\ref{net1}) of the form 
$\frac{P}{Q}dt-dx^{(n-1)}=0$ and adding null terms 
$S_1(t,x,x^{(1)},...,x^{(n-1)})$ $x^{(1)}dt-S_1(t,x,x^{(1)},...,x^{(n-1)})dx $ and
$S_i(t,x,x^{(1)},...,x^{(n-1)})x^{(i)}dt-S_i(t,x,x^{(1)},...,x^{(n-1)})$ $dx^{(i-1)}
,\;i=2,3,...n-1$,  
to it we obtain that on the solutions the 1-form
\begin{eqnarray}
\bigg(\frac{P}{Q}+\sum_{i=1}^{n-1}S_i x^{(i)}\bigg)dt-S_1dx
-\sum_{i=2}^{n-1}S_i dx^{(i-1)}-dx^{(n-1)}= 0. 
\label{net6} 
\end{eqnarray}	
Hence, on the solutions, the 1-forms (\ref{net3}) and 
(\ref{net6}) must be proportional. Multiplying (\ref{net6}) by the 
factor $ R(t,x,x^{(1)},...,x^{(n-1)})$ which acts as the integrating factor
for (\ref{net6}), we have on the solutions that 
\begin{eqnarray} 
dI=R\bigg[\bigg(\phi+\sum_{i=1}^{n-1}S_i x^{(i)}\bigg)dt-S_1dx
-\sum_{i=2}^{n-1}S_idx^{(i-1)}-dx^{(n-1)}\bigg]= 0, 
\label{net7}
\end{eqnarray}
where $ \phi\equiv {P}/{Q}$. Comparing equations (\ref{net3}) 
with (\ref{net7}) we have, on the solutions, the relations 
\begin{eqnarray} 
I_{t}  = R(\phi+\sum_{i=1}^{n-1}S_i x^{(i)}),
\;\;\;\;\;\;\;\;\;\;\;\;\;\;\;\;\;\;\nonumber\\
I_{x}  = -RS_1,
\;\;\;\;\;\;\;\;\;\;\;\; \;\;\;\;\;\;\;\;\;\;\;\;\;\;\;\;\;\;\;\;\;\;\;\nonumber\\
I_{x^{(i)}}  = -RS_{i+1},\;\;i=1,2,\ldots,n-2, \;\nonumber\\
I_{x^{(n-1)}} = -R.
\;\;\;\;\;\;\;\;\;\;\;\;\;\;\;\;\;\;\;\;\;\;\;\;\;\;\;\;\;\;\;\;\;\;\;\;\;\;\;
 \label{net8}
\end{eqnarray} 
The compatibility conditions, 
$I_{tx}=I_{xt}$, $I_{tx^{(n-1)}}=I_{{x^{(n-1)}}t}$, $I_{xx^{(n-1)}}=I_{{x^{(n-1)}}x}$,
$I_{t{x^{(i)}}}=I_{{x^{(i)}}t}$, $I_{x{x^{(i)}}}=I_{{x^{(i)}}x}$,
$I_{{x^{(i)}}x^{(n-1)}}=I_{x^{(n-1)}{x^{(i)}}}$, $i=2,3,\ldots,n-2$, between the 
equations (\ref{net8}), provide us the conditions
\begin{eqnarray}
D[S_1] & = -\phi_x+S_1\phi_{x^{(n-1)}}+S_1S_{n-1},
\qquad \qquad \qquad \qquad \qquad \qquad\;\; \label{net9}\\
D[S_i] & = -\phi_{x^{(i-1)}}+S_i\phi_{x^{(n-1)}}+S_iS_{i+1}-S_{i-1},
\;i=2,3,\dots,n-2,\label{net10}\\
D[S_{n-1}]&=-\phi_{x^{(n-2)}}+S_{n-1}\phi_{x^{(n-1)}}+S_{n-1}^2-S_{n-2} 
, \qquad \qquad \qquad \quad \;\;\label{net11}\\
D[R]  &= -R(S_{n-1}+\phi_{x^{(n-1)}}),
\qquad \qquad \qquad \qquad \qquad \qquad \qquad \quad\;\;\label{net12}\\
R_{x^{(i)}}S_1 & = -R{S_{1}}_{x^i}+R_{x}S_{i+1}+R{S_{i+1}}_{x}, 
\;\;i=1,2,\dots,n-2,  \quad \qquad\;  \label{net13}\\
R_{x^{(i)}}S_{j+1} & = -R{S_{j+1}}_{x^{(i)}}+R_{x^j}S_{i+1}+R{S_{i+1}}_{x^j},
\;\;i,j=1,2,\dots,n-2,\;\; \label{net14}\\
R_{x^{(i)}} & = R_{x^{(n-1)}}S_{i+1}+R{S_{i+1}}_{x^{(n-1)}},
\;\;i=1,2,\dots,n-2, \qquad \qquad \;\;\label{net15}\\
R_x & = R_{x^{(n-1)}}S_1+R{S_{1}}_{x^{(n-1)}},
\qquad \qquad \qquad \qquad \qquad \qquad \qquad \quad \label{net16}
\end{eqnarray}
where the total differential operator $D$ is defined by 
\begin{eqnarray}
D=\frac{\partial}{\partial{t}}+
x^{(1)}\frac{\partial}{\partial{x}}
+\sum_{i=2}^{n}x^{(i)}\frac{\partial}{\partial{x^{(i-1)}}}.
\nonumber
\end{eqnarray}

We note that (\ref{net9})-(\ref{net16}) form an overdetermined system for
the unknowns, $S_i,\;i=1,\dots,n-1$, and $R$. For example, one can check that 
for a
second order ODE, $n=2$, one gets three equations for two unknowns, say, $S$ and
$R$, whereas for a third order ODE one gets six equations for
three unknowns, say, $S_1$, $S_2$ and $R$. Thus in this procedure, for a given $n^{th}$ order
ODE, one gets $\frac{n(n+1)}{2}$ number of equations, for $n$ unknowns, out of
which $\frac{n(n-1)}{2}$ equations are just extra constraints.

The crux of the problem lies in solving the determining equations and
identifying sufficient number of integrating factors and null forms. 
 But the point is that any particular solution will suffice for the purpose. 
 We solve equations~(\ref{net9})-(\ref{net16}) 
in the following way.  
Substituting the expression for $\phi=\frac{P}{Q}$, obtained from 
equation~(\ref{net1}), into (\ref{net9})-(\ref{net11}) 
we get a system of differential equations for the unknowns $S_i,\;i=1,\dots,n-1$. 
Solving 
them we can obtain expressions for  the null forms $S_i$'s.  Once $S_i$'s are 
known then  
equation~(\ref{net12}) becomes the determining equation for the function $R$.  
Solving the latter we can get an explicit form for $R$.  
Now the functions $R$ and $S_i,\;i=1,\dots,n-1$, have to satisfy the extra constraints  
(\ref{net13})-(\ref{net16}).    
However, once a compatible solution satisfying all the equations have 
been found then the functions $R$ and $S_i,\;i=1,2,\dots,n-1$, fix the first 
integral $I(t,x,x^{(1)},...,x^{(n-1)})$ by the relation 
\begin{eqnarray}
I(t,x,x^{(1)},...,x^{(n-1)}) = \sum_{i=1}^{n}r_i-\int\left[R+\frac{d}{dx^{(n-1)}} 
\left(\sum_{i=1}^{n}r_i
   \right)\right]dx^{(n-1)},
  \label{net17}
\end{eqnarray}
where 
\begin{eqnarray} 
r_1 & = \int R\bigg(\phi+\sum_{i=1}^{n-1}S_i x^{(i)}\bigg)dt,
\qquad \qquad \qquad \qquad \qquad \quad \quad \;\;\nonumber\\
r_2 &=-\int \bigg(RS_1+\frac{d}{dx}r_1\bigg) dx,
\qquad \qquad \qquad \qquad \qquad \quad \quad \qquad \;\;\;\nonumber\\
r_j &=-\int \left[RS_{j-1}+\frac{d}{dx^{(j-1)}}
\left(\sum_{k=1}^{j-1}r_k\right)\right]dx^{(j-1)},\;\;j=3,\ldots,n. \nonumber
\end{eqnarray}
Equation~(\ref{net17}) can be derived straightforwardly by integrating the 
equations~(\ref{net8}). Now substituting the expressions for $\phi$, $R$ and
$S_i,i=1,2,\dots,n$,
into (\ref{net17}) and evaluating the integrals one can get the associated
integrals of motion. However, we have to point out that we have not examined the
question of existence of consistent solutions to 
equations~(\ref{net9})-(\ref{net16}) at present.
\section{Method of generating integrals of motion}
\label{sec3}
In the above, we derived the integrals of motion, $I_i,\; i=1,2,\ldots,n$, 
by constructing sufficient
number of integrating factors. Interestingly, for a {\it class of equations},
one can also generate the required
number of integrals of motion from one of the integrals, say $I_1$, if it is
known, provided its form can be written in a specific form. For example, for
the $n^{th}$ order equation~(\ref{net1}), one can generate
$I_2,I_3,\ldots,I_{n-1}$ and
$I_{n}$ from $I_1$ itself. In the following we illustrate this possibility.

Let us assume that there exists a first integral for the $n^{th}$ order equation 
(\ref{net1}) of the form, $I_1(t,x,x^{(1)},x^{(2)},...,x^{(n-1)})=C$. Now 
let us split the 
functional form of the first integral $I_1$ into two terms such that 
one involves all the variables 
$(t,x,x^{(1)},...,x^{(n-1)})$ while the other excludes $x^{(n-1)}$, that is,
\begin{eqnarray}  
I_1=F_1(t,x,x^{(1)},...,x^{(n-1)})+F_2(t,x,x^{(1)},...,x^{(n-2)}). \label{the01}
\end{eqnarray}
Of course, such a splitting is not unique, which can in fact be used profitably
further to identify new integrals.

Now let us split 
the function $F_1$ further in terms of two functions such that $F_1$ itself 
is a function of the product of the two functions, say, a perfect 
differentiable function  $\frac {d}{dt}G_1(t,x,x^{(1)},...,x^{(n-2)})$ and 
another function  $G_2(t,x,x^{(1)},...,x^{(n-1)})$, that is,
\begin{eqnarray} 
I_1=F_1\left(\frac{1}{G_2(t,x,x^{(1)},...,x^{(n-1)})}
\frac{d}{dt}G_1(t,x,x^{(1)},...,x^{(n-2)})\right)\qquad \qquad\qquad\nonumber\\
\qquad  +F_2\left(G_1(t,x,x^{(1)},...,x^{(n-2)})\right).
\label{the02}
\end{eqnarray}
We note that while rewriting equation (\ref {the01}) in the form (\ref {the02}), we
demand the function $F_2(t,x,x^{(1)},...,x^{(n-2)})$ in (\ref {the01}) automatically 
to be a function of $G_1(t,x,x^{(1)},...,x^{(n-2)})$. Now identifying  the 
function  $G_1$ as the new dependent variable and the integral of $G_2$ over 
time as the new independent variable, that is,
\begin{eqnarray} 
w = G_1(t,x,x^{(1)},...,x^{(n-2)}),\quad 
z = \int_o^t G_2(t',x,x^{(1)},...,x^{(n-1)}) dt', 
\label{the03}
\end{eqnarray}
one can rewrite equation (\ref {the02}) of the form
\begin{eqnarray} 
I=F_1\left(\frac {dw}{dz}\right)+F_2(w).
\end{eqnarray}
In other words
\begin{eqnarray} 
F_1\left(\frac {dw}{dz}\right)=I-F_2(w).\label {eq13a}
\end{eqnarray}
Now rewriting equation (\ref {eq13a}) one obtains a separable equation 
\begin{eqnarray}  
\frac {dw}{dz}=f(w),\label {eq13b}
\end{eqnarray}
which can be integrated by quadrature, and $I_2$ can be obtained.

The procedure given above is easy to follow and can be used to solve 
{ \it a class of
problems} straightforwardly. In fact, for the linearizable second order ODEs, we
find that the function $F_2$ turns out to be zero and as a consequence one gets
$\frac{dw}{dz}=I_1$ (from (\ref {eq13b})) which can be integrated to obtain the
second integration constant trivially. On the other hand, if $F_2$ is not zero 
then the second
integration constant can be deduced after the integration of (\ref {eq13b}),
which can be done for a number of examples. As far as third order ODEs are
concerned the first order equation (\ref {eq13b}) provides us the second
integral whereas a different choice $\hat{G_1}$ and $\hat{G_2}$, where $\hat{G_1}$ and 
$\hat{G_2}$ are different from $G_1$ and $G_2$, directly leads us to the third
integral (see Sec.~5). The procedure can in principle be extended to higher
order ODEs.
\section{Second order ODEs}
\label{sec4}
The theory developed in Secs.~2 and 3, in principle, can be used to solve 
\textit{a class of
equations}. To illustrate the underlying ideas let us first consider a second order
ODE. Since we are going to discuss in detail only second and third order ODEs,
hereafter, we use the notation $\dot{x}$, $\ddot{x}$ and $\dddot{x}$ instead 
of $x^{(1)}$, $x^{(2)}$ and $x^{(3)}$ for $\frac{dx}{dt}$, $\frac{d^2x}{dt^2}$ 
and $\frac{d^3x}{dt^3}$, respectively. 

Fixing, $n=2$, the determining equations (\ref{net9})-(\ref{net16}) get 
simplified to  
\begin{align}  
S_t+\dot{x}S_x+\phi S_{\dot{x}} & = -\phi_x+S\phi_{\dot{x}}+S^2,\label{smet9}\\
R_t+\dot{x}R_x+\phi R_{\dot{x}}  & = -R(S+\phi_{\dot{x}}),\label{smet10}\\
R_x & = R_{\dot{x}}S+RS_{\dot{x}},
\label{smet11}
\end{align}
The integral of motion, (\ref{net17}), is fixed by the relation
\begin{eqnarray}
 I(t,x,\dot{x})= r_1
  -r_2 -\int \left[R+\frac{d}{d\dot{x}} \left(r_1-r_2\right)\right]d\dot{x},
  \label{smet13}
\end{eqnarray}
where 
\begin{eqnarray} 
r_1 = \int R(\phi+\dot{x}S)dt \quad \mbox{and} \quad
r_2 =\int (RS+\frac{d}{dx}r_1) dx. \nonumber
\end{eqnarray}
The method of solving the determining equations~(\ref{smet9})-(\ref{smet11}) is
described in detail in Ref.\cite{Duarte, Chand1}. However, in order to be self-contained, we
briefly summarize the main ideas in the following.

Let us first solve the equation~(\ref{smet9}) with the given $\phi$ and obtain
expression for $S$. Once $S$ is known then  
equation~(\ref{smet10}) becomes the determining equation for the function $R$.  
Solving the latter one can get an explicit form for $R$.  
Now the functions $R$ and $S$ have to satisfy an extra constraint, that is, 
equation~(\ref{smet11}).  We note at this point that all solutions which satisfy 
equations~(\ref{smet9})-(\ref{smet10}) need not satisfy the constraint 
(\ref{smet11}) 
since equations~(\ref{smet9})-(\ref{smet11}) constitute an overdetermined system 
for the unknowns $R$ and $S$. For example, let us consider a set $(S,R)$ be a 
solution of equations~(\ref{smet9})-(\ref{smet10}) and not of the constraint 
equation 
(\ref{smet11}).  After examining several examples we observe that one can make 
the set $(S,R)$ compatible by modifying the form of $R$ as
\begin{align}
\hat{R} & = F(t,x,\dot{x})R,
\label{smet101}
\end{align} 
where $\hat{R}$ satisfies equation~(\ref{smet10}), so that we have
\begin{eqnarray} 
(F_t+\dot{x}F_x+\phi F_{\dot{x}}) R+FD[R]
   =-FR(S+\phi_{\dot{x}}).
\label{smet102}
\end{eqnarray}
Further, if $F$ is a constant of motion (or a function of it), then the first 
term on the left hand side
vanishes and one gets the same equation~(\ref{smet10}) for $R$ provided $F$ is 
non-zero.
In other words, whenever $F$ is a non-zero constant or a function of the 
integral of motion then the solution of
equation~(\ref{smet10}) may provide only a factor of the complete solution 
$\hat{R}$ without the
factor $F$ in equations~(\ref{smet101}).  This general form of $\hat{R}$ with $S$ 
will form a complete
solution to the equations~(\ref{smet9})-(\ref{smet11}).  In a nutshell we 
describe the
procedure as follows.  First we determine $S$ and $R$ from 
equations~(\ref{smet9})-(\ref{smet10}). If the set $(S,R)$ satisfies 
equation~(\ref{smet11})
then we take it as a compatible solution.  On the other hand if it does not
satisfy (\ref{smet11}) then we assume the modified form $\hat{R}=F(I)R$, and 
find the explicit form of $F(I)$  from equation~(\ref{smet11}), which in turn 
fixes the compatible solution $(S,\hat{R})$.
Once $I_1$ is derived then the second integration constant can be
deduced either utilizing our procedure described in Sec.~3, or finding another
set of solutions $(S_2,R_2)$ for equations~(\ref{smet9})-(\ref{smet11}). The
method has been applied to several interesting nonlinear dynamical systems and
interesting results have been obtained \cite{Chand1}.
In the following example we illustrate both the above ideas.
\subsection{Example }
Let us consider an equation of the following form for illustrative purpose:
\begin{eqnarray}            
\ddot{x}=\frac{(2x-1)}{(1+x^2)}{\dot{x}}^2
\label{scat101}
\end{eqnarray}
so that the equations~(\ref{smet9})-(\ref{smet11}) become 
\begin{align}           
S_t+\dot{x}S_x+\frac{(2x-1){\dot{x}}^2}{(1+x^2)} S_{\dot{x}} & = 
2\bigg(\frac{x(1+x)-1}{(1+x^2)^2}\bigg){\dot{x}}^2
+\frac{2(2x-1)\dot{x}}{(1+x^2)}S+S^2,
\label{scat102}\\
R_t+\dot{x}R_x+\frac{(2x-1){\dot{x}}^2}{(1+x^2)} R_{\dot{x}} & =
-\bigg(S+\frac{2(2x-1)\dot{x}}{(1+x^2)}\bigg)R,
\label{scat103}\\
R_x-SR_{\dot{x}}-RS_{\dot{x}} & = 0.
\label{scat104}                                                         
\end{align} 
As mentioned in Sec. 2, any particular solution satisfying equations
(\ref{scat102})-(\ref{scat104}) is sufficient to derive an integral of motion.
 We solve equations
(\ref{scat102})-(\ref{scat104}) in the following way. Equation (\ref{scat102}) 
is a first order partial differential equation in $S$
with variable coefficients. To seek a particular solution for $S$ we consider a
simple ansatz for $S$ of the form 
\begin{eqnarray} 
S = {a(t,x)+b(t,x)\dot{x}},
\label{cat15}
\end{eqnarray}
where $a$ and $b$ are arbitrary functions of $t$ and $x$ (in other examples one
may need to take rational forms in $\dot{x}$). 
Substituting (\ref{cat15}) into (\ref{scat102}) and equating the coefficients 
of different powers of $\dot{x}$ to zero we get a set of partial differential 
equations for the variables $a$ and $b$.  Solving them we arrive at
\begin{equation}  
S_1 = \frac{(2x-1)}{(1+x^2)}\dot{x},  \quad 
S_2 = -\frac{1}{t}+\frac{(1-2x)}{(1+x^2)}\dot{x}.
\label{scat105}
\end{equation}
Substituting the forms of $S_1$ and $S_2$ separately into (\ref{scat103}) and solving 
the resultant equations 
one can obtain the corresponding forms of $R$. Let us first 
consider $S_1$.  Substituting the latter into (\ref{scat103}) we obtain an
equation for $R$ of the form
\begin{eqnarray}           
R_t+\dot{x}R_x+\frac{(2x-1){\dot{x}}^2}{(1+x^2)} R_{\dot{x}} & =
-\bigg(\frac{(2x-1)\dot{x}}{(1+x^2)}\bigg)R.
\label{scat103a}                                                       
\end{eqnarray} 
One can immediately identify a particular solution 
\begin{equation}
R_1= -\frac{e^{tan^{-1}x}}{(1+x^2)},
\label{scat106}
\end{equation}
to this equation with a polynamial ansatz in $\dot{x}$. One can easily check 
that $S_1$ and $R_1$ satisfy equation~(\ref{scat104}) also. As a consequence 
one can deduce the first integral, using the relation (\ref{smet13}), of the 
form
\begin{eqnarray}     
I_1 =\frac{\dot{x}e^{tan^{-1}x}}{(1+x^2)}.  
\label{scat107}
\end{eqnarray}

Now substituting the expression $S_2$ into 
(\ref{scat103}) we obtain an equation for $R$. As in the previous case, one can
easily fix a particular solution of the form 
\begin{equation} 
R_{2}=-\frac{t}{\dot{x}}.
\label{scat108}
\end{equation}
However, this set $(S_2,R_2)$ does not satisfy the extra constraint 
(\ref{scat104}) and so to deduce the correct form of $R_2$ we assume that 
\begin{align}
\hat{R}_2 & = F(I_1)R_2 = 
-F(I_1)\frac{t}{\dot{x}},
\label{scat109}
\end{align}
where $F$ is an arbitrary function of the first integral $I_1$. Substituting 
(\ref{scat109}) into equation~(\ref{scat104}) we obtain 
$F = {\frac {1}{I_1}}$, which fixes the form of $\hat{R}$ as 
\begin{eqnarray}         
\hat{R}_2 = -\frac{te^{tan^{-1}x}}{(1+x^2)}. 
\label{scat110}  
\end{eqnarray}
Now one can easily check that this set $(S_2,\hat{R}_2)$ is a compatible 
solution for the set (\ref{scat102})-(\ref{scat104}) which in turn provides 
$I_2$ through the relation (\ref{smet13}) in the form
\begin{eqnarray} 
I_2=e^{tan^{-1}x}\bigg(1-\frac{t\dot{x}}{(1+x^2)}\bigg).
\label{scat111}
\end{eqnarray}
Using the explicit forms of the first integrals $I_1$ and $I_2$, the solution 
of Eq.~(\ref{smet9}) can be deduced directly as
\begin{eqnarray}
x(t)=\tan\bigg[\log{(I_1t+I_2)}\bigg].
\label{scat112}
\end{eqnarray}

However, as shown in the Sec.~3, one can also deduce the second integral 
from the first integral itself. By using the procedure indicated in Sec. 3, by demanding that $I_1$
be put in the form
\begin{eqnarray} 
I_1=F_1\left(\frac{1}{G_2(t,x,\dot{x})}\frac{d}{dt}G_1(t,x)\right)
+F_2\left(G_1(t,x)\right),
\label{1smet13b}
\end{eqnarray}
so that in the transformed variables become
$w = G_1(t,x)$ and $z = \int_o^t G_2(t',x,\dot{x}) dt'$ and a first order ODE
results which can be solved by quadrature.
For example, in the present case, it is easy to rewrite the first integral 
(\ref{scat107}) in the form (\ref{1smet13b}), by inspection, namely,
\begin{eqnarray}
I_1=\frac {d}{dt}\left(e^{tan^{-1}x}\right),
\label {gr01} 
\end{eqnarray}
and identifying (\ref{gr01}) with (\ref{the02}), we get
\begin{eqnarray}
G_1=e^{tan^{-1}x},\;\;\;\;\; 
G_2=1,\;\;\;\;\;  F_2=0.
\label {gr02} 
\end{eqnarray}
With the above choices, equation~(\ref{the03}) furnishes the transformation
variables,
\begin{eqnarray}
w=e^{tan^{-1}x},\;\;\;\;\;  
z=t.
\label {gr03}  
\end{eqnarray}
Substituting (\ref{gr03}) into (\ref{gr01}) we get
\begin{eqnarray} 
\frac {dw}{dz}=I_1,
\end{eqnarray}
which in turn gives the free particle equation by differentiation or leads to 
the solution (\ref{scat112}) by an integration.  On the other hand vanishing of 
the function $F_2$ in this analysis is precisely the condition for the system 
to be transformed into the free particle equation.
\section{Third order ODEs}
\label{sec5}
Now we focus our attention on third order ODEs, $n=3$ in equation~(\ref{net1}).
Fixing $n=3$ in the determining equations~(\ref{net9})-(\ref{net16}), we get 
\begin{eqnarray}
D[S_1] & = -\phi_x+S_1\phi_{\ddot{x}}+S_1S_2,
\;\;\;\;\;\;\label{met9}\\
D[S_2] & = -\phi_{\dot{x}}+S_2\phi_{\ddot{x}}+S_2^2-S_1,\;\label{met10}\\
D[R]  &= -R(S_2+\phi_{\ddot{x}}),
\;\;\;\;\;\;\;\;\;\;\;\;\;\;\;\;\label{met11}\\
R_{\dot{x}}S_1 & = -R{S_1}_{\dot{x}}+R_{x}S_2+R{S_2}_{x}, \label{met12}\\
R_{\dot{x}} & = R_{\ddot{x}}S_2+R{S_2}_{\ddot{x}},
\;\;\;\;\;\;\;\;\;\;\;\;\;\;\;\; \label{met13}\\
R_x & = R_{\ddot{x}}S_1+R{S_1}_{\ddot{x}},
\;\;\;\;\;\;\;\;\;\;\;\;\;\;\;\;\label{met14}
\end{eqnarray}
where the total differential operator $D$ is defined by 
\begin{eqnarray}
D=\frac{\partial}{\partial{t}}+
\dot{x}\frac{\partial}{\partial{x}}+\ddot{x}\frac{\partial}{\partial{\dot{x}}}
+\phi\frac{\partial}{\partial{\ddot{x}}}.
\nonumber
\end{eqnarray}
The associated integral of motion is fixed by the relation
\begin{eqnarray}
I = r_1-r_2-r_3-\int\left(R+\frac{d}{d\ddot{x}} \left( r_1-r_2-r_3
   \right)\right)d\ddot{x},
  \label{met15}
\end{eqnarray}
where 
\begin{eqnarray} 
r_1&= \int R(\phi+S_1\dot{x}+S_2\ddot{x})dt,\qquad \qquad
r_2&=\int ( RS_1+\frac{d}{dx}r_1) dx,\nonumber\\
r_3&=\int \left(RS_2+\frac{d}{d\dot{x}} 
\left(r_1-r_2\right)\right)d\dot{x}. \qquad \quad\;\;\nonumber
\end{eqnarray}
As mentioned earlier, (\ref{met9})-(\ref{met14}) form an overdetermined system for
the unknowns, $S_1$, $S_2$ and $R$. We solve the (\ref{met9})-(\ref{met14}) 
in the following way.  
Substituting the expression for $\phi$ into (\ref{met9})-(\ref{met10}) 
we get a system of differential equations for the unknowns $S_1$ and $S_2$. 
Solving 
them we can obtain expressions for  the null forms $(S_1,S_2)$.  Once $S_2$ is 
known then  
equation~(\ref{met11}) becomes the determining equation for the function $R$.  
Solving the latter we can get an explicit form for $R$.  
Now the functions $R,S_1$ and $S_2$ have to satisfy the extra constraints  
(\ref{met12})-(\ref{met14}). 
\subsection{Example: 1}   
Let us consider an equation discussed by Bluman and Anco from the symmentres
point of view \cite{Bluman},
\begin{eqnarray}            
\dddot{x}=\frac{6t{\ddot{x}}^3}{{\dot{x}}^2}+\frac{6{\ddot{x}}^2}{\dot{x}}.
\label{cat101}
\end{eqnarray}
Substituting $\phi =\frac{6t{\ddot{x}}^3}{{\dot{x}}^2}+\frac{6{\ddot{x}}^2}{\dot{x}}$ 
into (\ref{met9})-(\ref{met11}), we get 
\begin{eqnarray}
{S_1}_t+\dot{x}{S_1}_x+\ddot{x}{S_1}_{\dot{x}}
+\bigg(\frac{6t{\ddot{x}}^3}{{\dot{x}}^2}+\frac{6{\ddot{x}}^2}{\dot{x}}\bigg)
{S_1}_{\ddot{x}} &= S_1\bigg(\frac{18t{\ddot{x}}^2}{{\dot{x}}^2}
 +\frac{12\ddot{x}}{\dot{x}}+S_2\bigg),\qquad \qquad \qquad \qquad 
\nonumber\\
{S_2}_t+\dot{x}{S_2}_x+\ddot{x}{S_2}_{\dot{x}}
+\bigg(\frac{6t{\ddot{x}}^3}{{\dot{x}}^2}+\frac{6{\ddot{x}}^2}{\dot{x}}\bigg)
{S_2}_{\ddot{x}} 
 &= \frac{12t\ddot{x}}{{\dot{x}}^3}+\frac{6{\ddot{x}}^2}{{\dot{x}}^2}
 +S_2\bigg(\frac{18t{\ddot{x}}^2}{\dot{x}^2}+\frac{12\ddot{x}}{\dot{x}}\bigg)
 +{S_2}^2-S_1,\nonumber\\
R_t+\dot{x}R_x+\ddot{x}R_{\dot{x}}
+\bigg(\frac{6t{\ddot{x}}^3}{{\dot{x}}^2}+\frac{6{\ddot{x}}^2}{\dot{x}}\bigg)
R_{\ddot{x}}&= -R\bigg(S_2+\frac{18t{\ddot{x}}^2}{{\dot{x}}^2}
 +\frac{12\ddot{x}}{\dot{x}}\bigg).\qquad \qquad \qquad \qquad \nonumber\\
 \label{cat104}
\end{eqnarray}
One can easily verify that equation~(\ref{cat104}) admits the following 
solutions, 
\begin{eqnarray}
S_{1}=0,\quad  S_{2}=-\frac{6t{\ddot{x}}^2+3\dot{x}\ddot{x}}{{\dot{x}}^2},
\quad R=\frac{{\dot{x}}^3}{{\ddot{x}}^2},\\
\bar{S_{1}}=0, \quad \bar{S_{2}}=-\frac{6t{\ddot{x}}^2+4\dot{x}\ddot{x}}{{\dot{x}}^2},
\quad \bar{R}=\frac{{\dot{x}}^4}{{\ddot{x}}^2},\\ 
\hat{S_{1}}=\frac{2{\ddot{x}}^2}{{\dot{x}}^2}, \quad 
\hat{S_{2}}=-\frac{6t{\ddot{x}}^2+2\dot{x}\ddot{x}}{{\dot{x}}^2}, \quad 
\hat{R}=\frac{{\dot{x}}^2}{{\ddot{x}}^2}. 
\label{cat105}
\end{eqnarray}
Having determined the functions $S_i$'s and $R$, $i=1,2$, one can proceed to
determine the associated integrals of motion. Substituting the expressions
into (\ref{met15}) separately and evaluating the
integrals, one obtains    
\begin{eqnarray} 
I_1 &=3t{\dot{x}}^2+\frac{{\dot{x}}^3}{\ddot{x}},\qquad \label{cat111}\\
I_2 &=2t{\dot{x}}^3+\frac{{\dot{x}}^4}{\ddot{x}},\qquad\label{cat112}\\
I_3 & =2x-6t\dot{x}-\frac{{\dot{x}}^2}{\ddot{x}},\; \label{cat114}
\end{eqnarray}
respectively.
One can easily check that $I_i's , i=1,2,3$, are constants on the solutions, 
that is, $\frac{dI_i}{dt} = 0, i=1,2,3$. 
From the integrals, $I_1$, $I_2$ and $I_3$, we can deduce the general solution
for the equation (\ref{cat101}) straightforwardly.

In the following we generate second and third integrals, say, (\ref{cat112}) and
(\ref{cat114}), from the first integral, that is, (\ref{cat111}), using our
procedure described in Sec.~3. 

Rewriting (\ref{cat111}) in the form (\ref{the02}) we get
\begin{eqnarray}   
I_1& =\displaystyle{-\frac {1}{\ddot{x}}\frac {d}{dt}(-t{\dot{x}}^3)}
=\displaystyle{\frac {dt}{dz}\frac {dw}{dt} = \frac {dw}{dz}},\label{the11}
\end{eqnarray}
so that 
\begin{eqnarray}  
w=-t{\dot{x}}^3,\quad z=-\dot{x}. \label{the12}
\end{eqnarray}
Integrating (\ref{the11}) and rewriting the latter in terms of the old variables 
we get
\begin{eqnarray} 
I_2 =2t{\dot{x}}^3+\frac{{\dot{x}}^4}{\ddot{x}},\label{the14}
\end{eqnarray}
which is exactly the same as the one we derived (vide equation (\ref{cat112})) 
earlier through the PS procedure.

To generate $I_3$ from $I_1$ we rewrite the latter in the form (\ref{the02}) but
with different functions $\hat{w}$ and $\hat{z}$, namely,
\begin{eqnarray}  
I_1 =-\frac {{\dot{x}}^2}{\ddot{x}}\frac {d}{dt}(2x-3t\dot{x})
= \frac {dt}{d\hat{z}}\frac {d\hat{w}}{dt} = \frac {d\hat{w}}{d\hat{z}},
 \;\;\;\;\;\;\;\label{the15}
\end{eqnarray}
so that
\begin{eqnarray}  
\hat{w}=2x-3t\dot{x},\quad \hat{z}=\dot{x}. \label{the16}
\end{eqnarray}
Integrating (\ref{the15}) we get 
\begin{eqnarray} 
\hat{w}=I_1\hat{z}+I_3  \Rightarrow I_3=\hat{w}-I_1\hat{z}.\label{the17}
\end{eqnarray}
Substituting (\ref{the16}) and (\ref{cat111}) into (\ref{the17}) we get
\begin{eqnarray}     
I_3  =2x-6t\dot{x}-\frac{{\dot{x}}^2}{\ddot{x}}, \label{the18}
\end{eqnarray}
which exactly coincides with (\ref{cat114}).

In a similar way one can derive $I_1$ and $I_2$ from $I_3$ and $I_1$ and $I_3$
from $I_2$.
\subsection{Example: 2} 
 Let us consider another nontrivial example which was discussed by 
 Steeb \cite{Steeb} in the context of invertible point transformations, namely,
\begin{eqnarray} 
\dddot{x}+\frac{3\dot{x}\ddot{x}}{x}-3\ddot{x}-\frac{3\dot{x}^2}{x}+2\dot{x}=0. 
\label{fthe20}
\end{eqnarray}
Substituting the form 
$\phi =-\frac{3}{x}(\dot{x}\ddot{x}-x\ddot{x}-\dot{x}^2+\frac{2}{3}x\dot{x})$ 
into (\ref{met9})-(\ref{met11}) and solving them, we find the following three
particular solutions which satisfy the determining equations
\begin{eqnarray}
(S_1,F_1,R_1)& =(\frac{\ddot{x}-\dot{x}}{x},
\frac{2\dot{x}-x}{x},xe^{-2t}),\;\label{fthe21a}\\
(S_2,F_2,R_2)& =(\frac{\ddot{x}-2\dot{x}}{x},
\frac{2\dot{x}-2x}{x},xe^{-t}),\;\label{fthe21b}\\
(S_3,F_3,R_3)& =(\frac{\ddot{x}-3\dot{x}+2x}{x},
\frac{2\dot{x}^2-3x}{x},x).
\label{fthe21c}
\end{eqnarray} 
Substituting the expressions (\ref{fthe21a})-(\ref{fthe21c}) separately into 
the expression 
(\ref{met15}) and evaluating the integrals we obtain 
\begin{eqnarray} 
I_1 &=(\dot{x}^2+x\ddot{x}-x\dot{x})e^{-2t},\qquad \label{cat111a}\\
I_2 &=(\dot{x}^2+x\ddot{x}-2x\dot{x})e^{-t},\qquad\label{cat112a}\\
I_3 & =(\dot{x}^2+x\ddot{x}-3x\dot{x}+x^2),\;\;\; \label{fthe22}
\end{eqnarray}
from which the general solution can be written of the form
\begin{eqnarray} 
x(t)=\bigg(\frac{I_1}{2}e^{2t}+I_2e^{t}+I_3\bigg)^{\frac{1}{2}}. 
\label{fthe22a}
\end{eqnarray}
From the integrals $I_1, I_2$ and $I_3$, one can also obtain the linearizing
transformation.
\section{Extension to Higher order ODEs} 
In the previous two sections we discussed the PS procedure applicable for 
second and third
order ODEs. Following the same steps one can derive the determining equations
for the fourth order ODE also from the equations (\ref{net9})-(\ref{net16}). This can
be done by restricting to $n=4$ in these equations. For example, for the
present case one obtains ten equations for four unknowns, namely,
$S_1,S_2,S_3,$ and $R$. Solving them consistently one can obtain explicit
expressions for the null forms, $S_i$'s, $i=1,2,3,$ and the integrating factor $R$.
Once the $S_i$'s and $R$ are known then the associated integral of motion can be
constructed using the relation (\ref{net17}) with $n=4$. Here also, if one is
able to find four sets of particular solutions, say, $(S_{ij},R_i),\;i=1,2,3,4$
and $j=1,2,3$, one can straightforwardly construct four integrals of motion. On
the other hand if one has less number of integrals of motion then using the
procedure described in Sec. 3 one can generate the remaining integrals of motion
from the known ones and establish the integrability. The extension to higher
order ODEs follows along similar lines.
\section{ Extension to coupled second order ODEs}
So far we discussed the applicability of the PS procedure to scalar differential
equations. Interstingly, the procedure can also be extended to coupled
ODEs. In the following we describe the procedure to two coupled second
order ODEs and the application to higher dimensional equations will be
discussed elsewhere.

Let us consider a class of second order ODEs of the form
\begin{eqnarray} 
\ddot{x}=\frac{d^2x}{dt^2}=\frac{P_1}{Q_1}, \quad
\ddot{y}={\frac{d^2y}{dt^2}}={\frac{P_2}{Q_2}}, \quad
{ P_i,Q_i}\in C{[t,x,y,\dot{x},\dot{y}]},\; i=1,2.\label {cso01} 
\end{eqnarray}
Let us suppose that the system (\ref{cso01}) admits a first integral of 
the form $I(t,x,y,\dot{x},\dot{y})=C$
 with C constant on the solution, so that the total differential gives 
\begin{eqnarray}  
dI={I_t}{dt}+{I_{x}}{dx}+{I_{y}}{dy}+{I_{\dot{x}}}{d\dot{x}}
+{I_{\dot{y}}}{d\dot{y}}=0, \label {cso02} 
\end{eqnarray}
where subscript denotes partial differentiation with respect to that variable. 
Rewriting ({\ref{cso01}) in the form
\begin{eqnarray}
\frac{P_1}{Q_1}dt-d\dot{x}=0,\qquad
\frac{P_2}{Q_2}dt-d\dot{y}=0 \label {cso03}
\end{eqnarray}
and adding null terms $S_1(t,x,y,\dot{x},\dot{y})\dot{x}dt
-S_1(t,x,y,\dot{x},\dot{y})dx $ and
$S_2(t,x,y,\dot{x},\dot{y})\dot{y}dt\;\;\;\;\;$ $-
S_2(t,x,y,\dot{x},\dot{y})dy $ suitably,
 we obtain that, on the solutions, the 1-forms
\begin{subequations}
\begin{eqnarray}
(\frac{P_1}{Q_1}+S_1\dot{x})dt-S_1dx-d\dot{x}=0,\label {cso04}\\
(\frac{P_2}{Q_2}+S_2\dot{y})dt-S_2dy-d\dot{y}=0.\label {cso05}
\end{eqnarray}
\label {cso06}
\end{subequations}

Hence, on the solutions, the 1-forms (\ref{cso02}) and 
(\ref{cso06}) must be proportional. Multiplying (\ref{cso04}) by the factor 
$ R_1(t,x,y,\dot{x},\dot{y})$ and (\ref{cso05}) by
the factor $ R_2(t,x,y,\dot{x},\dot{y})$, which act as the integrating 
factors for (\ref{cso04}) and (\ref{cso05}), respectively, we have on the 
solutions that 	
\begin{eqnarray} 
dI=R_1(\phi_1+S_1\dot{x})dt+R_2(\phi_2+S_2\dot{y})dt-R_1S_1dx
-R_2S_2dy-R_1d\dot{x}-R_2d\dot{y}=0,\;\;\label {cso07}
\end{eqnarray}
where $ \phi_i\equiv {P_i}/{Q_i},\; i=1,2$. Comparing equations (\ref{cso07}) 
and (\ref{cso02}) we have, on the solutions, the relations 
\begin{subequations}
\begin{eqnarray} 
 I_t  =R_1(\phi_1+S_1\dot{x})+R_2(\phi_2+S_2\dot{y}), \;\;\;\;\\
 I_{x}  = -R_1S_1,\qquad \qquad \qquad \qquad \qquad \;\;\;\; \\
 I_{y} =-R_2S_2, \qquad \qquad \qquad \qquad \qquad \quad \\
 I_{\dot{x}}  =-R_1,\qquad \qquad \qquad \qquad \qquad \qquad  \\
 I_{\dot{y}} =-R_2.\qquad \qquad \qquad \qquad \qquad \qquad \;
\end{eqnarray}
\label {cso08}
\end{subequations}
The compatibility conditions between the equations 
(\ref{cso08}) provide us the conditions,
\begin{eqnarray} 
D{[S_1]} =-\phi_{1x}-\frac{R_2}{R_1} \phi_{2x}
	+\frac{R_2}{R_1}S_1\phi_{2\dot{x}}
           +S_1\phi_{1\dot{x}}+S_1^2,  \label {eq23}\\
D{[S_2]} =-\phi_{2y}-\frac{R_1}{R_2} \phi_{1y}
	+ \frac{R_1}{R_2}S_2\phi_{1\dot{y}}
           +S_2\phi_{2\dot{y}}+S_2^2,  \label {eq24}\\
D{[R_1]}  =-{(R_1\phi_{1\dot{x}}+R_2\phi_{2\dot{x}}+R_1 S_1)}, 
\qquad \qquad \qquad  \label {eq25}\\
D{[R_2]}  =-{(R_2\phi_{2\dot{y}}+R_1\phi_{1\dot{y}}+R_2 S_2)}, 
\qquad \qquad \qquad  \label {eq26}\\
S_1R_{1y} =-R_1S_{1y}+S_2R_{2x}+R_2S_{2x}, 
\qquad \qquad \qquad \;  \label {eq31}\\
R_{1x} =S_1R_{1\dot{x}}+R_1S_{1\dot{x}},
\qquad \qquad \qquad \qquad \qquad \quad  \label {eq27}\\
R_{2y} =S_2R_{2\dot{y}}+R_2S_{2\dot{y}}, 
\qquad \qquad \qquad \qquad \qquad \quad \label {eq28}\\   
R_{1y} =S_2R_{2\dot{x}}+R_2S_{2\dot{x}}, 
\qquad \qquad \qquad \qquad \qquad \quad  \label {eq29}\\  
R_{2x} =S_1R_{1\dot{y}}+R_1S_{1\dot{y}},  
\qquad \qquad \qquad \qquad \qquad \quad   \label {eq30}\\
R_{1\dot{y}} =R_{2\dot{x}}, 
\qquad \qquad \qquad \qquad \qquad \qquad \qquad \qquad   \label {eq32}
\end{eqnarray}
where the total differential operator is now defined by
\begin{eqnarray} 
D &=\frac{\partial}{\partial{t}}+\dot{x}\frac{\partial}{\partial{x}}
+\dot{y}\frac{\partial}{\partial{y}}
+\phi_1\frac{\partial}{\partial{\dot{x}}}+\phi_2\frac{\partial}
{\partial{\dot{y}}}. &
\end{eqnarray}
Integrating equations~(\ref{cso08}), we obtain the integral of motion,
\begin{eqnarray}
I=r_1+r_2+r_3+r_4
-\int\bigg[R_2+\frac{d}{d\dot{y}}\bigg(r_1+r_2+r_3+r_4\bigg)\bigg]
d\dot{y},
\label {cso09}
\end{eqnarray}
where
\begin{eqnarray}
r_1&=\int\bigg(R_1(\phi_1+S_1\dot{x})+R_2(\phi_2+S_2\dot{y})\bigg)dt,
\qquad 
r_2=-\int\bigg(R_1S_1+\frac{d}{dx}(r_1)\bigg)dx,&\qquad\nonumber\\
r_3&=-\int\bigg(R_2S_2+\frac{d}{dy}(r_1+r_2)\bigg)dy,\qquad 
r_4=-\int\bigg[R_1+\frac{d}{d\dot{x}}\bigg(r_1+r_2+r_3\bigg)\bigg]d\dot{x}.&
\nonumber
\end{eqnarray}

As we did earlier, solving the determining equations 
(\ref{eq23})-(\ref{eq32}) consistently we can obtain expressions for the 
function $S_i$'s and $R_i$'s, $i=1,2$.
Substituting them into (\ref{cso09}) and evaluating the integrals we can deduce
the associated integrals of motion.
However, unlike the scalar case, the determining equations in the present case are highly
coupled and pose difficulties to approach them directly. To overcome this 
problem we adopt the following technique.
We rewrite  equations~(\ref{eq23})-(\ref{eq32}) for two variables, namely, 
$R_1$ and
$R_2$, by eliminating $S_1$ and $S_2$, and solve the resultant equations and
obtain expressions for $R_1$ and $R_2$. From the latter we deduce the forms of
$S_1$ and $S_2$ by using the relations (\ref{eq25}) and (\ref{eq26}). To
implement this algorithm we use all the equations~(\ref{eq23})-(\ref{eq32})
effectively such that the functions $S_i$'s $i=1,2$, can be eliminated and 
the functions $R_1$ and $R_2$ can be deduced from an optimal set of equations.

To begin with we deduce the following two identities,
\begin{eqnarray} 
D{[R_1 S_1]} &=-(R_1\phi_{1x}+R_2\phi_{2x}), & \label {eq33}\\
D{[R_2S_2]} &=-(R_1\phi_{1y}+R_2\phi_{2y}), & \label {eq34}
\end{eqnarray}
which can be obtained by combining (\ref {eq23})-(\ref {eq26}). Now we have
explicit forms for the total derivatives $R_1S_1$ and $R_2S_2$. Let us now take
total derivative of  equations~(\ref{eq25}) and (\ref{eq26}) 
and substitute (\ref{eq33}) and (\ref{eq34}) in
the resultant equations to obtain
\begin{eqnarray}
R_{1tt}+2\dot{x}R_{1tx}+2\dot{y}R_{1ty}+2\phi_1R_{1t\dot{x}}
+2\phi_2R_{1t\dot{y}}+\dot{x}^2R_{1xx}
+2\dot{x}\dot{y}R_{1xy}+\dot{y}^2R_{1yy}+\phi_{1t}R_{1\dot{x}}\nonumber\\
+\phi_{2t}R_{1\dot{y}}+\dot{x}\phi_{1x}R_{1\dot{x}}
+\dot{y}\phi_{1y}R_{1\dot{x}}+2\dot{x}\phi_1 R_{1x\dot{x}}
+2\dot{y}\phi_1 R_{1y\dot{x}}+2\dot{x}\phi_2 R_{1x\dot{y}}
+2\dot{y}\phi_2 R_{1y\dot{y}}\nonumber\\
+\dot{x}\phi_{2x}R_{1\dot{y}}+\dot{y}\phi_{2y}R_{1\dot{y}}
+\phi_1R_{1x}+\phi_2R_{1y}+\phi_1\phi_{1\dot{x}}R_{1\dot{x}}
+\phi_1^2R_{1\dot{x}\dot{x}}+\phi_2\phi_{1\dot{y}}R_{1\dot{x}}\nonumber\\
+\phi_1\phi_{2\dot{x}}R_{1\dot{y}}+\phi_2\phi_{2\dot{y}}R_{1\dot{y}}
+\phi_{1\dot{x}}(R_{1t}+\dot{x}R_{1x}+\dot{y}R_{1y}+\phi_1R_{1\dot{x}}
+\phi_2R_{1\dot{y}})\nonumber\\
+\phi_2^2R_{1\dot{y}\dot{y}}+2\phi_1\phi_2R_{1\dot{x}\dot{y}}
+\phi_{2\dot{x}}(R_{2t}+\dot{x}R_{2x}+\dot{y}R_{2y}
+\phi_1R_{2\dot{x}}+\phi_2R_{2\dot{y}})\nonumber\\
-R_1\phi_{1x}-R_2\phi_{2x}
+R_1(\phi_{1t\dot{x}}+\dot{x}\phi_{1x\dot{x}}+\dot{y}\phi_{1y\dot{x}}
+\phi_1\phi_{1\dot{x}\dot{x}}+\phi_2\phi_{1\dot{x}\dot{y}})\nonumber\\
+R_2(\phi_{2t\dot{x}}+\dot{x}\phi_{2x\dot{x}}+\dot{y}\phi_{2y\dot{x}}
+\phi_1\phi_{2\dot{x}\dot{x}}+\phi_2\phi_{2\dot{x}\dot{y}})=0,\nonumber\\ 
\label {eq35}\\
R_{2tt}+2\dot{x}R_{2tx}+2\dot{y}R_{2ty}+2\phi_1R_{2t\dot{x}}
+2\phi_2R_{2t\dot{y}}+\dot{x}^2R_{2xx}
+2\dot{x}\dot{y}R_{2xy}+\dot{y}^2R_{2yy}+\phi_{1t}R_{2\dot{x}}\nonumber\\
+\phi_{2t}R_{2\dot{y}}+\dot{x}\phi_{1x}R_{2\dot{x}}
+\dot{y}\phi_{1y}R_{2\dot{x}}+2\dot{x}\phi_1 R_{2x\dot{x}}
+2\dot{y}\phi_1 R_{2y\dot{x}}+2\dot{x}\phi_2 R_{2x\dot{y}}
+2\dot{y}\phi_2 R_{2y\dot{y}}\nonumber\\
+\dot{x}\phi_{2x}R_{2\dot{y}}+\dot{y}\phi_{2y}R_{2\dot{y}}
+\phi_1R_{2x}+\phi_2R_{2y}+\phi_1\phi_{1\dot{x}}R_{2\dot{x}}
+\phi_1^2R_{2\dot{x}\dot{x}}+\phi_2\phi_{1\dot{y}}R_{2\dot{x}}\nonumber\\
+\phi_1\phi_{2\dot{x}}R_{2\dot{y}}+\phi_2\phi_{2\dot{y}}R_{2\dot{y}}
+\phi_{1\dot{y}}(R_{1t}+\dot{x}R_{1x}+\dot{y}R_{1y}+\phi_1R_{1\dot{x}}
+\phi_2R_{1\dot{y}})\nonumber\\
+\phi_2^2R_{2\dot{y}\dot{y}}+2\phi_1\phi_2R_{2\dot{x}\dot{y}}
+\phi_{2\dot{y}}(R_{2t}+\dot{x}R_{2x}+\dot{y}R_{2y}
+\phi_1R_{2\dot{x}}+\phi_2R_{2\dot{y}})\nonumber\\
-R_1\phi_{1y}-R_2\phi_{2y}
+R_1(\phi_{1t\dot{y}}+\dot{x}\phi_{1x\dot{y}}+\dot{y}\phi_{1y\dot{y}}
+\phi_1\phi_{1\dot{x}\dot{y}}+\phi_2\phi_{1\dot{y}\dot{y}})\nonumber\\
+R_2(\phi_{2t\dot{y}}+\dot{x}\phi_{2x\dot{y}}+\dot{y}\phi_{2y\dot{y}}
+\phi_1\phi_{2\dot{x}\dot{y}}+\phi_2\phi_{2\dot{y}\dot{y}})=0.\nonumber\\ 
\label {eq35a}
\end{eqnarray}
equations~(\ref{eq27})-(\ref{eq32}) can also be written of the form
\begin{eqnarray} 
 R_{1x}=\frac{\partial}{\partial{\dot{x}}}(R_1S_1),\;\;
 R_{2y}=\frac{\partial}{\partial{\dot{y}}}(R_2S_2),\;\;
 R_{1y}=\frac{\partial}{\partial{\dot{x}}}(R_2S_2),\;\;
 R_{2x}=\frac{\partial}{\partial{\dot{y}}}(R_1S_1).
 \label {sec01}
\end{eqnarray}
These identities help us to obtain some additional equations for the variables
$R_1$ and $R_2$. For example
differentiating (\ref {eq25}) with respect to $ \dot{x}$ and (\ref {eq26})
with respect to $\dot{y}$ and using the identities (\ref {sec01}) 
in the resulting equations, we get
\begin{eqnarray} 
 R_{1t\dot{x}}+\dot{x}R_{1x\dot{x}}+\dot{y}R_{1y\dot{x}}
 +\phi_1 R_{1\dot{x}\dot{x}} +\phi_2 R_{2\dot{x}\dot{x}}+2R_{1x}
+2\phi_{2\dot{x}}R_{2\dot{x}}+2\phi_{1\dot{x}}R_{1\dot{x}}\nonumber\\
+R_2\phi_{2\dot{x}\dot{x}}+R_1\phi_{1\dot{x}\dot{x}}=0, 
\label {eq36}\\
R_{2t\dot{y}}+\dot{x}R_{2x\dot{y}}+\dot{y}R_{2y\dot{y}}
+\phi_1 R_{1\dot{y}\dot{y}}+\phi_2 R_{2\dot{y}\dot{y}}+2R_{2y}
+2\phi_{2\dot{y}}R_{2\dot{y}}+2\phi_{1\dot{y}}R_{1\dot{y}}\nonumber\\
+R_2\phi_{2\dot{y}\dot{y}}+R_1\phi_{1\dot{y}\dot{y}}=0. \label {eq37}
\end{eqnarray}
On the other hand, differentiating (\ref {eq25}) with respect to $\dot{y}$ and
using  (\ref {sec01}) we get 
\begin{eqnarray}
R_{1t\dot{y}}+\dot{x}R_{1x\dot{y}}+\dot{y}R_{1y\dot{y}}
+\phi_1 R_{1\dot{x}\dot{y}}+\phi_2 R_{1\dot{y}\dot{y}}+R_{1y}
+R_{2x}+\phi_{2\dot{x}}R_{2\dot{y}}\nonumber\\+\phi_{1\dot{x}}R_{1\dot{y}}
+R_2\phi_{2\dot{x}\dot{y}}+R_1\phi_{1\dot{x}\dot{y}}
+\phi_{1\dot{y}}R_{1\dot{x}}+\phi_{2\dot{y}}R_{1\dot{y}}=0. \label {eq38}
\end{eqnarray}
The remaining possibility, that is, differentiation of (\ref {eq26}) 
with respect to $\dot{x}$, leads us to the same equation (\ref {eq38}) and so we
can discard it.

We can further simplify equations~(\ref {eq35}) and (\ref {eq35a}) by utilizing the
equations~(\ref {eq36})-(\ref {eq38}). The final form of the
equations~(\ref {eq35}) and (\ref {eq35a}) reads
\begin{eqnarray}
R_{1tt}+2\dot{x}R_{1tx}+2\dot{y}R_{1ty}+\phi_1R_{1t\dot{x}}
+\phi_2R_{1t\dot{y}}+\dot{x}^2R_{1xx}+2\dot{x}\dot{y}R_{1xy}
-R_1\phi_{1x}\nonumber\\
-R_2\phi_{2x}+\dot{y}^2R_{1yy}
+\phi_{1t}R_{1\dot{x}}+\phi_{2t}R_{1\dot{y}}
+\dot{x}\phi_{1x}R_{1\dot{x}}+\dot{y}\phi_{1y}R_{1\dot{x}}
+\dot{x}\phi_1 R_{1x\dot{x}}\nonumber\\+\dot{y}\phi_1 R_{1y\dot{x}}
+\dot{x}\phi_2 R_{1x\dot{y}}
+\dot{y}\phi_2 R_{1y\dot{y}}+\dot{x}\phi_{2x}R_{1\dot{y}}
+\dot{y}\phi_{2y}R_{1\dot{y}}
-\phi_1R_{1x}\nonumber\\-\phi_2R_{2x}
+\phi_{1\dot{x}}(R_{1t}+\dot{x}R_{1x}+\dot{y}R_{1y})
+R_1(\phi_{1t\dot{x}}+\dot{x}\phi_{1x\dot{x}}+\dot{y}\phi_{1y\dot{x}})\nonumber\\
+\phi_{2\dot{x}}(R_{2t}+\dot{x}R_{2x}+\dot{y}R_{2y})
+R_2(\phi_{2t\dot{x}}+\dot{x}\phi_{2x\dot{x}}+\dot{y}\phi_{2y\dot{x}})=0, 
\label {eq39}\\\nonumber\\
R_{2tt}+2\dot{x}R_{2tx}+2\dot{y}R_{2ty}+\phi_1R_{2t\dot{x}}
+\phi_2R_{2t\dot{y}}+\dot{x}^2R_{2xx}+2\dot{x}\dot{y}R_{2xy}
-R_1\phi_{1y}\nonumber\\-R_2\phi_{2y}
+\phi_{1t}R_{2\dot{x}}+\phi_{2t}R_{2\dot{y}}+\dot{y}^2R_{2yy}
+\dot{x}\phi_{1x}R_{2\dot{x}}+\dot{y}\phi_{1y}R_{2\dot{x}}
+\dot{x}\phi_1 R_{2x\dot{x}}\nonumber\\+\dot{y}\phi_1 R_{2y\dot{x}}
+\dot{x}\phi_2 R_{2x\dot{y}}+\dot{y}\phi_2 R_{2y\dot{y}}
+\dot{x}\phi_{2x}R_{2\dot{y}}+\dot{y}\phi_{2y}R_{2\dot{y}}
-\phi_1R_{1y}\nonumber\\-\phi_2R_{2y}
+\phi_{1\dot{y}}(R_{1t}+\dot{x}R_{1x}+\dot{y}R_{1y})
+R_1(\phi_{1t\dot{y}}+\dot{x}\phi_{1x\dot{y}}+\dot{y}\phi_{1y\dot{y}})
\nonumber\\
+\phi_{2\dot{y}}(R_{2t}+\dot{x}R_{2x}+\dot{y}R_{2y})
+R_2(\phi_{2t\dot{y}}+\dot{x}\phi_{2x\dot{y}}+\dot{y}\phi_{2y\dot{y}})=0. 
\label {eq40}
\end{eqnarray}
As a result now we have a system of five equations (of course in second order) 
for the
unknowns $R_1$ and $R_2$, namely equations~(\ref{eq36})-(\ref{eq40}). 
Substituting the expressions for $\phi_1$ and $\phi_2$ 
into (\ref{eq36})-(\ref{eq40}) and solving them one gets the integrating factors 
$R_1$and $R_2$. Once $R_i$'s, $i=1,2,$ are known the null forms $S_i$'s, 
$i=1,2,$ can be fixed through the relation (\ref{eq25})-(\ref{eq26}). 
\subsection{Example: Two-dimensional Kepler problem}
Here we consider a simple, but physically important example, namely, two
dimensional Kepler problem and illustrate the method developed in the
previous section.

Let us consider the Kepler problem in the $x-y$ plane, that is,
\begin{eqnarray} 
\ddot{\hat{{\bf{r}}}}+\frac{\hat{{\bf{r}}}}{r^3}=0,
 \label {kep01}
\end{eqnarray}
where $\hat{{\bf{r}}}=x\hat{{\bf{i}}}+y\hat{{\bf{j}}}$ and $r=|\hat{\bf{r}}|$. The respective equations of
motions are
\begin{eqnarray}
\ddot{x}=-\frac {x}{(x^2+y^2)^{\frac {3}{2}}}=\phi_1(x,y), \\
\ddot{y}=-\frac {y}{(x^2+y^2)^{\frac {3}{2}}}=\phi_2(x,y).  \label {kep02}
\end{eqnarray}
In this case the determining
equations~(\ref{eq36})-(\ref{eq40}) simplify to
\begin{eqnarray} 
 R_{1t\dot{x}}+\dot{x}R_{1x\dot{x}}+\dot{y}R_{1y\dot{x}}+\phi_1 R_{1\dot{x}\dot{x}}
 +\phi_2 R_{2\dot{x}\dot{x}}+2R_{1x}=0, 
 \label {eq42}\\
R_{2t\dot{y}}+\dot{x}R_{2x\dot{y}}+\dot{y}R_{2y\dot{y}}+\phi_1 R_{1\dot{y}\dot{y}}
+\phi_2 R_{2\dot{y}\dot{y}}+2R_{2y}=0, 
\label {eq43}\\
R_{1t\dot{y}}+\dot{x}R_{1x\dot{y}}+\dot{y}R_{1yv{y}}+\phi_1 R_{1\dot{x}\dot{y}}
+\phi_2 R_{1\dot{y}\dot{y}}+R_{1y}+R_{2x}=0,
\label {eq44}
\end{eqnarray}
\begin{eqnarray}
R_{1tt}+2\dot{x}R_{1tx}+2\dot{y}R_{1ty}+\dot{x}^2R_{1xx}+2\dot{x}\dot{y}R_{1xy}
+\dot{y}^2R_{1yy}+\dot{x}\phi_{1x}R_{1\dot{x}}+\dot{y}\phi_{1y}R_{1\dot{x}}
\nonumber\\+\dot{x}\phi_1 R_{1x\dot{x}}
+\dot{y}\phi_1 R_{1y\dot{x}}
+\dot{x}\phi_2 R_{1x\dot{y}}+\dot{y}\phi_2 R_{1y\dot{y}}+\dot{x}\phi_{2x}R_{1\dot{y}}
+\dot{y}\phi_{2y}R_{1\dot{y}}\nonumber\\
+\phi_1R_{1t\dot{x}}+\phi_2R_{1t\dot{y}}-R_1\phi_{1x}-R_2\phi_{2x}
-\phi_1R_{1x}-\phi_2R_{2x}=0, 
\label {eq45}\\
R_{2tt}+2\dot{x}R_{2tx}+2\dot{y}R_{2ty}+\dot{x}^2R_{2xx}+2\dot{x}\dot{y}R_{2xy}
+\dot{y}^2R_{2yy}+\dot{x}\phi_{1x}R_{2\dot{x}}+\dot{y}\phi_{1y}R_{2\dot{x}}
\nonumber\\+\dot{x}\phi_1 R_{2x\dot{x}}
+\dot{y}\phi_1 R_{2y\dot{x}}+\dot{x}\phi_2 R_{2x\dot{y}}
+\dot{y}\phi_2 R_{2y\dot{y}}+\dot{x}\phi_{2x}R_{2\dot{y}}
+\dot{y}\phi_{2y}R_{2\dot{y}}\nonumber\\
+\phi_1R_{2t\dot{x}}+\phi_2R_{2t\dot{y}}-R_1\phi_{1y}-R_2\phi_{2y}
-\phi_1R_{1y}-\phi_2R_{2y}=0. 
\label {eq46}
\end{eqnarray}
To solve equations~(\ref{eq42})-(\ref{eq46}) we seek an ansatz
\begin{eqnarray}
R_1=a_1(x,y)+a_2(x,y)\dot{x}+a_3(x,y)\dot{y},\label {eq48}\\
R_2=b_1(x,y)+b_2(x,y)\dot{x}+b_3(x,y)\dot{y}, \;\;\label {eq48a}
\end{eqnarray}
where $a_i$'s and $b_i$'s, $i=1,2,3,$ are arbitrary functions of $x$ and $y$. 
Substituting (\ref {eq48}) and (\ref {eq48a}) into (\ref{eq42})-(\ref{eq46}) and
solving the resultant equations we can obtain at least the following three
solutions. 
\begin{eqnarray}
(i)\; \quad R_1=\dot{x},\qquad  \qquad \;\; R_2=\dot{y},\qquad \label {kep03}\\
(ii)\;\quad \bar{R_1}=y,\qquad \qquad \;\;  \bar{R_2}=x,\qquad\label {kep04}\\
(iii)\;\quad \hat{R_1}=2y\dot{x}-x\dot{y},\;\quad \hat{R_2}=-x\dot{x}.\;\;\;\label {kep05}
\end{eqnarray}
We are also now searching for other possible forms with a modified ansatz. 
Substituting (\ref {kep03}) into (\ref{eq25}) and (\ref{eq26}) we get
\begin{eqnarray}
(i)\;\quad S_1=\displaystyle{\frac {x}{\dot{x}(x^2+y^2)^{\frac {3}{2}}}},
\quad S_2=\frac {y}{\dot{y}(x^2+y^2)^{\frac {3}{2}}}. \label {kep06}
\end{eqnarray}
In similar way equations (\ref {kep04}) and (\ref {kep05}) provide us
\begin{eqnarray}
(ii)\;\quad\bar{S_1}=-\frac{\dot{y}}{y},\;\;\quad\qquad \qquad 
\bar{S_2}=-\frac{\dot{x}}{x},\qquad \quad\; \label{kep07}\\
(iii)\;\quad\hat{S_1}=-\frac{(-\dot{x}\dot{y}+\frac {xy}{r^3})}{2y\dot{x}-x\dot{y}},
\quad 
\hat{S_2}=\frac{(\dot{x}^2-\frac {x^2}{r^3})}{x\dot{x}}.\quad \label{kep08}
\end{eqnarray}
Substituting the expressions $R_i$'s and $S_i$'s, $i=1,2$, into (\ref {cso09})
and evaluating the integrals we get,
\begin{eqnarray}
I_1=\frac {1}{2}(\dot{x}^2+\dot{y}^2)-\frac{1}{\sqrt{x^2+y^2}}, \label {kep09}
\end{eqnarray}
which is of course the Hamiltonian of the system. The forms $\bar{R_i}$'s and 
$\bar{S_i}$'s, $i=1,2$, provide us
\begin{eqnarray}
I_2=y\dot{x}-x\dot{y}, \label {kep10}
\end{eqnarray}
the second integral, namely, the angular momentum. The integrating factors 
$\hat{R_i}$'s and
null form $\hat{S_i}$'s, $i=1,2$, lead us to
\begin{eqnarray}
I_{3}={\dot{x}(y\dot{x}-x\dot{y})-\frac {y}{\sqrt{x^2+y^2}}},& \label {kep11}
\end{eqnarray}
namely, the Runge-Lenz constant.
\section{Final Remarks}
In this paper, we have discussed the method of solving a class of ODEs through
the modified PS method. The method is applicable to both scalar and
multicomponent equations of any order. We also demonstrated the theory with
examples. Apart from the above, in the scalar case, we introduced a novel way of generating integral
of motion from a single integral and illustrated our ideas with the same
example considered previously. The application of this method to multicomponent
systems and their integrability and linearization properties will be published
elsewhere.

\section*{Acknowledgements}
The work of VKC is supported by Council of Scientific and Industrial Research,
India. The work of MS and ML forms part of a 
Department of Science and Technology, Government of India, sponsored 
research project.

\label{lastpage}

\end{document}